\documentclass[aps,reprint,amsmath]{revtex4-1}
\usepackage{graphicx}
\usepackage{dsfont}
\usepackage{amsfonts}

\newcommand{\bra}[1]{\langle #1 | \,}
\newcommand{\ket}[1]{\, | #1 \rangle}
\newcommand{\braket}[2]{\langle #1 | #2 \rangle}
\newcommand{\expv}[1]{\langle #1 \rangle}
\newcommand{\ga}{\gamma}
\newcommand{\Ga}{\Gamma}

\newcommand{\de}{\delta}
\newcommand{\De}{\Delta}

\newcommand{\Om}{\Omega}
\newcommand{\hlf}{\frac{1}{2}}
\newcommand{\lra}{\leftrightarrow}
\newcommand{\sig}{\hat{\sigma}}
\newcommand{\Sig}{\hat{\Sigma}}
\newcommand{\hL}{\hat{L}}
\newcommand{\mc}[1]{\mathcal{#1}}

\begin{document}

\title{On the adiabatic preparation of spatially-ordered Rydberg excitations of atoms \\
in a one-dimensional optical lattice by laser frequency sweeps}

\author{David Petrosyan}
%\email{david.petrosyan@iesl.forth.gr}
\affiliation{Institute of Electronic Structure and Laser, FORTH, 
GR-71110 Heraklion, Crete, Greece}

\author{Klaus M\o lmer}
\affiliation{Department of Physics and Astronomy, Aarhus University,
DK-8000 Aarhus C, Denmark}

\author{Michael Fleischhauer}
\affiliation{Fachbereich Physik und Forschungszentrum OPTIMAS, 
Technische Universit\"at Kaiserslautern, D-67663 Kaiserslautern, Germany}

\date{\today}

\begin{abstract}
We examine the adiabatic preparation of crystalline phases of Rydberg 
excitations in a  one-dimensional lattice gas by frequency sweep 
of the excitation laser, as proposed by 
Pohl \textit{et al.} [Phys. Rev. Lett. \textbf{104}, 043002 (2010)] 
and recently realized experimentally by 
Schau{\ss} \textit{et al.} [Science \textbf{347}, 1455 (2015)].
We find that the preparation of crystals of a few Rydberg excitations 
in a unitary system of several tens of atoms requires exceedingly long 
times for the adiabatic following of the ground state of the system Hamiltonian.
Using quantum stochastic (Monte-Carlo) wavefunction simulations, 
we show that realistic decay and dephasing processes affecting 
the atoms during the preparation lead to a final state of the system 
that has only a small overlap with the target crystalline state. 
Yet, the final number and highly sub-Poissonian statistics of Rydberg 
excitations and their spatial order are little affected by the relaxations.
\end{abstract}

%\pacs{ }

\maketitle

\section{Introduction}

Atoms in high-lying Rydberg states strongly interact with each other via the 
long-range dipole-dipole or van der Waals potentials \cite{rydQIrev,rydDBrev}. 
These interactions can suppress multiple Rydberg excitations of atoms within 
a certain blockade distance from each other \cite{Lukin2001}, and are being 
explored for obtaining ordered phases of interacting many-body systems
and simulating quantum phase transitions \cite{Weimer2008,Weimer10,Sela2011,%
Zeller2012,Lesanovsky2011,Lesanovsky2012,Lee2011,Hoening2013,Hoening2014,Ji2011}.

Several conceptually different approaches have been suggested for 
the preparation of crystalline order of Rydberg excitations in 
spatially-extended ensembles of cold atoms. These include direct 
(near-)resonant laser excitation of strongly-interacting Rydberg 
states in continuous or lattice gases
\cite{Ates2012Ji2013,Garttner2012,DPMHMF2013,Petrosyan2013,Schauss2012,Labuhn2015},
and the deceleration and storage in one-dimensional (1D) atomic medium 
of propagating light pulses forming the so-called Rydberg polaritons 
under the conditions of electromagnetically induced transparency 
\cite{Otterbach2013,Maxwell2013}.
A common feature of all these schemes is that the resulting spatially-periodic
structure of Rydberg excitations has finite density-density correlation length, 
% -- typically of the order of the Rydberg blockade distance -- 
while their number exhibits highly sub-Poissonian statistics characterized 
by negative Mandel parameter $Q \lesssim -0.5$. The Rydberg excitations
then essentially form a liquid rather than a crystal phase with long-range 
order. 

To achieve perfect Rydberg crystals with long -- ideally infinite -- 
correlation length, an adiabatic preparation protocol of the ground
state of an Ising-type Hamiltonian for interacting Rydberg gases has 
been proposed \cite{Pohl2010,Schachenmayer2010,vanBijnen2011,Vermersch2015}. 
An experimental realization involving a few Rydberg excitations in a 1D lattice 
of several tens of sites was recently reported in Ref.~\cite{Schauss2015}.
Our aim here is to critically re-examine this protocol, taking into 
account realistic relaxation processes affecting the atoms. We find that, 
under typical experimental conditions, it is not feasible to attain the
perfectly-ordered ground state of the Hamiltonian even for three or four 
Rydberg excitations in a finite 1D lattice gas. This is because the atomic 
decay and dephasing during the exceedingly long preparation time required 
for the adiabatic evolution of the system spoil the adiabaticity and 
significantly reduce the overlap of the final state of the system with 
the target ground state of the Hamiltonian. This overlap, or fidelity, 
is largest at some intermediate value of the preparation time, and 
maximizing the probability of the ordered state of Rydberg excitations 
requires therefore a compromise between the adiabatic following and 
decoherence. Even though the perfectly-ordered state cannot be obtained with 
high fidelity, good spatial ordering of Rydberg excitations is still achieved. 
%The corresponding mixed state is, however, very close to the steady state 
%of a continuously-driven dissipative system \cite{DPMHMF2013,Ates2012Ji2013}.

\section{The adiabatic preparation protocol}

We consider a system of $N$ atoms trapped in a 1D optical lattice potential, 
with one atom per lattice site, assuming no site-occupation defects.
A spatially-uniform time-dependent laser field couples the ground 
state $\ket{g}$ of each atom to the Rydberg state $\ket{r}$ with 
the Rabi frequency $\Om(t)$ and detuning $\de(t) \equiv \omega-\omega_{rg}$. 
In the frame rotating with the laser field frequency $\omega$, 
the system is described by the Ising--spin-$\hlf$--like Hamiltonian 
\begin{equation}
\mc{H}/\hbar = - \de(t) \sum_{j}^N \sig_{rr}^j 
+ \sum_{i<j}^N \De_{ij} \sig_{rr}^i \sig_{rr}^j
- \Om(t) \sum_{j}^N (\sig_{rg}^j + \sig_{gr}^j) ,  \label{eq:Ham}
\end{equation}
where $\sig_{\mu \nu}^j \equiv \ket{\mu}_{jj}\bra{\nu}$ 
are the projection ($\mu = \nu$) or transition ($\mu \neq \nu$) operators 
for atom $j$, while $\De_{ij} = C_6/r_{ij}^6$ is the strength of the 
(repulsive, $C_6 >0$) van der Waals interaction between the Rydberg-excited 
atoms $i$ and $j$ separated by distance $r_{ij}$. 

%%%%%%%%%%%%%%%%% FIGURE %%%%%%%%%%%%%%%%%%
\begin{figure}[t]
\includegraphics[width=8.7cm]{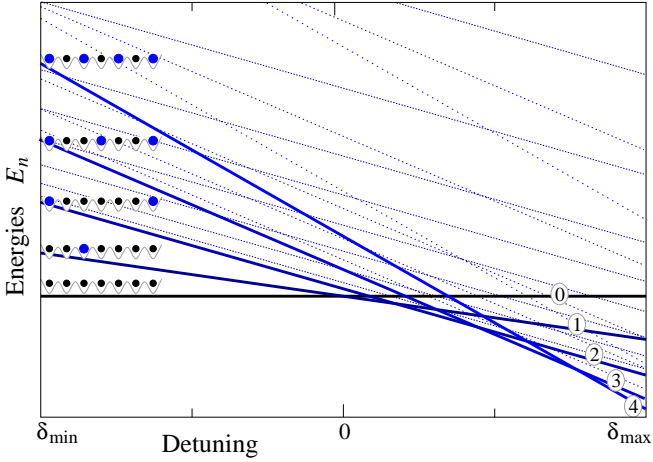}
\caption{Diagram of energies $E_n$ of Hamiltonian~(\ref{eq:Ham}) in the 
limit of $\Om \to 0$ versus laser detuning $\de$, for $n=0,1,2,3,4$ Rydberg
excitations of atoms in a lattice. Thick lines correspond to the lowest 
energy $E_n^{\min}$ within the $n$-excitation subspace, while thin dotted 
lines with the same slope (and color) denote the excited state energies
with the same $n \geq 2$.}  
\label{fig:sysEs}
\end{figure}
%%%%%%%%%%%%%%%%%%%%%%%%%%%%%%%%%%%

Our aim is to prepare the ground state of Hamiltonian~(\ref{eq:Ham}) 
in the classical limit of $\Om \to 0$.
The complete basis consists of states with $n=0,1,2,\ldots,N$ Rydberg 
excitations. On an $N$-site lattice, $n$ excitations can be arranged 
in $\binom{N}{n}$ different ways, which is the dimension of the 
corresponding subspace $\mathbb{H}_n$ of the total Hilbert space 
$\mathbb{H} = \sum_{n=0}^N \mathbb{H}_n$. In the absence of interactions,
$\De_{ij} = 0 \, \forall \, i,j \in [1,N]$, all the states in each 
$\mathbb{H}_n$ are degenerate, having the energy $E_n = - n \de$. 
Interactions $\De_{ij} > 0$ between the atoms (partially) lift this degeneracy,
unless $n =0$ with $E_0 = 0$ for the zero-excitation 
state $\ket{R_0} \equiv \ket{g g \ldots g}$, 
or $n = 1$ with $E_1 = - \de$ for all $N$ single-excitation states
and their symmetric superposition
$\ket{R_1} \equiv \frac{1}{\sqrt{N}} \sum_j \ket{gg \ldots r_j \ldots g}$.  
For $n \geq 2$, the lowest energy states $\ket{R_{n}^{\min}}$ are the states 
with the largest separation between the Rydberg excitations: 
$E_{2}^{\min} = -2 \de + \frac{C_6}{l^6}, \;
E_{3}^{\min} = -3 \de + \frac{C_6}{l^6} + 2 \frac{C_6}{(l/2)^6}, \;  
E_{4}^{\min} = -3 \de + \frac{C_6}{l^6} + 2 \frac{C_6}{(2l/3)^6} 
+ 3 \frac{C_6}{(l/3)^6}, \ldots $, and more generally 
\begin{eqnarray}
E_{n}^{\min} &=& -n \de + \frac{C_6 (n-1)^6}{l^6} \sum_{k=1}^{n-1} \frac{k}{(n-k)^6} 
\nonumber \\
&\simeq & -n \de + \frac{C_6 (n-1)^7}{l^6} ,
\end{eqnarray}
where $l = a (N-1)$ is the length of the system, i.e., the distance 
between the first and last atoms, with $a$ the lattice constant.  
The energy spectrum $\{ E_n \}$ versus detuning $\de$ is schematically 
shown in Fig.~\ref{fig:sysEs}. For negative detunings $\de = \de_0$, 
the ground state of the system corresponds to the $n=0$ excitation 
state $\ket{R_0}$ with $E_0 = 0$, while for positive $\de \simeq \de_n$ 
the ground state corresponds to the lowest-energy $n$ excitation state 
$\ket{R_{n}^{\min}}$, such that $E_{n}^{\min} < E_{n \pm 1}^{\min}$ which leads 
to $\de_n \simeq \frac{C_6 n^7}{2l^6}$.

In the adiabatic preparation protocol
\cite{Pohl2010,Schachenmayer2010,vanBijnen2011,Vermersch2015,Schauss2015}, 
we start with the state $\ket{R_0}$ and the laser detuning 
having some negative value $\de <0$ which we then slowly increase 
till reaching some positive final value $\de \simeq \de_n$.
As we vary the detuning, the energies $E_{0,1,\ldots,n}^{\min }$ cross 
at around $\de_{0\to 1} = 0, \; \de_{1 \to 2} = \frac{C_6}{l^6} , \;  
\de_{2 \to 3} \simeq \frac{C_6 2^7}{l^6}, \; \ldots \; ,
\de_{(n-1) \to n} \simeq \frac{C_6 [(n-1)^7-(n-2)^7]}{l^6}$ \cite{Pohl2010}.
Of course, with vanishing field amplitude $\Om \to 0$, there is no 
coupling and thereby transitions between the energy levels $E_n$, 
and the system initially in state $\ket{R_0} = \ket{gg\ldots g}$ 
will remain in that state irrespective of $\de$. Hence, as we change 
the detuning, the field $\Om$ should be non-zero when the energy 
levels $E_{0,1,\ldots,n}^{\min }$ cross, which would lead to avoided 
crossings and adiabatic following of the ground state of the system. 
The initial state with zero Rydberg excitations $\ket{R_0}$ is coupled 
by the field to the symmetric single excitation state $\ket{R_1}$ 
with the collectively-enhanced rate 
\[ \Om_{0\to 1} = \sqrt{N} \Om , \] 
which leads to a large level repulsion $\pm \Om_{0\to 1}$ 
in the vicinity of $\de_{0\to 1}$. 
Next, state $\ket{R_1}$ is coupled to the lowest-energy double-excitation
state $\ket{R_2^{\min}} \equiv \ket{r_1 g \ldots gr_N}$ with a much smaller 
rate of 
\[ \Om_{1 \to 2} = \frac{2 \Om}{\sqrt{N}} , \]
and the corresponding level repulsion around $\de_{1\to 2}$ is small $\pm \Om_{1\to 2}$.
In turn, state $\ket{R_2^{\min}}$ is coupled to the lowest-energy triple-excitation 
state $\ket{R_3^{\min}} \equiv \ket{r_1 g \ldots gr_{(N+1)/2} g \ldots  gr_N}$
(assuming odd $N$) with the single-atom transition rate 
\[ \Om_{2 \to 3} = \Om , \] 
and the levels are repelled by $\pm \Om_{2\to 3}$ around $\de_{2\to 3}$.
The lowest-energy four-excitation state $\ket{R_4^{\min}} 
\equiv \ket{r_1g \ldots gr_{(N+2)/3}g \ldots gr_{(2N+1)/3}g \ldots gr_N}$
(assuming $N= 6k +1$ with $k=1,2,3,\ldots$) is coupled to the three-excitation 
state $\ket{R_3^{\min}}$ only via three-photon process, and thus the transition 
amplitude is small \cite{Pohl2010}, 
\[ \Om_{3\to 4} \propto \frac{\Om^3}{ [C_6/(l/3)^6]^2 } . \] 
The successive transitions to higher $n \geq 4$ excitation states involve 
$(2n-5)$-photon processes yielding therefore even smaller
transition amplitudes $\Om_{(n-1) \to n}$. 

As we change the laser detuning, our intention
\cite{Pohl2010,Schachenmayer2010,vanBijnen2011,Schauss2015} is that, 
in the vicinity of $\de_{(n-1) \to n}$, the system adiabatically follows 
the ground state $\ket{R_{n-1}^{\min}} \to \ket{R_{n}^{\min}}$. 
According to the Landau-Zener formula \cite{LandauZener,Brundobler1993}, 
the probability of non-adiabatic transition 
$\ket{R_{n-1}^{\min}} \to \ket{R_{n-1}^{\min}}$
is given by $p_{\mathrm{n.a.}} \sim \exp(-2 \pi |\Om_{(n-1) \to n}|^2/\alpha)$, 
where $\alpha = \frac{\partial}{\partial t} |E_{n-1}^{\min } -E_{n}^{\min }|
= \frac{\partial}{\partial t} \de$ is determined by the rate of 
change of detuning $\de$. Hence, due to the small values of the
effective Rabi frequencies $\Om_{(n-1) \to n}$ for $n \geq 4$,
adiabatic population of states beyond $n =3$ will be difficult to achieve.
We therefore mainly focus on preparing adiabatically the triple-excitation 
state $\ket{R_{3}^{\min}}$, but we will briefly consider longer lattices 
which can accommodate $n =4$ excitations under otherwise similar conditions. 
Notice also the bottleneck for the transition $\ket{R_{1}} \to \ket{R_{2}^{\min}}$,
due to the smallness of the effective Franck-Condon factor ($f = 2/\sqrt{N}$) 
of $\Om_{1 \to 2}$, as compared to $\Om_{0 \to 1}$ ($f=\sqrt{N}$) and 
$\Om_{2 \to 3}$ ($f=1$). As will be illustrated below, this fact has rather 
interesting implications for the adiabatic preparation of the target 
double- and triple-excitation states $\ket{R_{2}^{\min}}$ and $\ket{R_{3}^{\min}}$.

We shall consider unitary dynamics of the system, as well as the influence
of relaxation processes. These include spontaneous decay of atoms 
from the Rydberg state $\ket{r}$ to the ground state $\ket{g}$ with 
rate $\Gamma_r$, and dephasing of the atomic transition $\ket{g} \lra \ket{r}$
with rate $\Gamma_z$ due to non-radiative collisions with the background 
atoms, external and trapping field noise, Doppler broadening, laser linewidth, 
and decay of the intermediate atomic state $\ket{e}$ to the ground 
state $\ket{g}$ when $\ket{g} \lra \ket{r}$ is a two-photon transition 
\cite{Schauss2012,Schauss2015}. The corresponding Lindblad generators 
for the decay and dephasing processes are 
$\hL_r^j = \sqrt{\Ga_r} \sig^j_{gr}$ and 
$\hL_z^j = \sqrt{\Ga_z} (\sig^j_{rr} -\sig^j_{gg})$, and the total decay 
rate of coherence $\expv{\sig_{rg}}$ on the transition $\ket{g} \lra \ket{r}$
is then $\ga_{rg} \equiv \hlf \Ga_{r} + 2 \Ga_{z}$ \cite{PLDP2007,Petrosyan2013}.
We assume closed systems in which 
$\sig_{gg}^j + \sig_{rr}^j = \mathds{1}^j \; \forall \; j \in [1,N]$
is preserved throughout evolution.
 
To simulate the dissipative dynamics of the many-body system, we employ 
the quantum stochastic (Monte Carlo) wavefunctions \cite{qjumps}.
In each quantum trajectory, the state of the system $\ket{\Psi(t)}$ 
evolves according to the Schr\"odinger equation 
$\partial_t \ket{\Psi} = -\frac{i}{\hbar} \tilde{\mc{H}} \ket{\Psi}$
with an effective Hamiltonian 
\begin{equation}
\tilde{\mc{H}} = \mc{H} - \frac{i}{2} \hbar \hL^2 ,
\end{equation}
where 
\[
\hL^2 \equiv \sum_j (\hL_r^{j\dagger} \hL_r^j + \hL_z^{j\dagger} \hL_z^j )
= \sum_j (\Ga_r \sig^j_{rr} + \Ga_z \mathds{1}^j)
\]
is the non-Hermitian part which does not preserve the norm of $\ket{\Psi}$ 
during the evolution. The evolution is interrupted by random quantum 
jumps $\ket{\Psi} \to \hL_{r,z}^j \ket{\Psi}$ with probabilities 
$W_{r,z}^j \equiv \bra{\bar\Psi} \hL_{r,z}^{j\dagger} \hL_{r,z}^j \ket{\bar\Psi}$,
where the normalized wavefunction of the system at any time $t$ is given by
$\ket{\bar{\Psi}(t)} = \ket{\Psi(t)}/\sqrt{\braket{\Psi(t)}{\Psi(t)}}$.
The expectation value of any observable $\hat{\mc{O}}$ of the system
is then obtained by averaging over many, $M \gg 1$, independently simulated
trajectories, $\expv{\hat{\mc{O}}} = \mathrm{Tr} [\hat{\rho}\hat{\mc{O}}]
= \frac{1}{M} \sum_m^M \bra{\bar{\Psi}_m} \hat{\mc{O}} \ket{\bar{\Psi}_m}$,
while the density operator is given by
$\hat{\rho}(t) = \frac{1}{M} \sum_m^M \ket{\bar{\Psi}_m(t)}
\bra{\bar{\Psi}_m(t)}$.

We define the populations of the lowest-energy $n$-excitation states as 
$P_n^{\mathrm{\min}} \equiv \bra{R_n^{\mathrm{\min}}}\hat{\rho}\ket{R_n^{\mathrm{\min}}}$.
%which, for a single trajectory, reduce to 
%$P_n^{\mathrm{\min}} \equiv |\braket{R_n^{\mathrm{\min}}}{\Psi}|^2$.
The mean number of Rydberg excitations within an ensemble of $N$ atoms
is $\expv{n} = \expv{\sum_j^N \sig_{rr}^j}$, while the probabilities 
$p_n = \expv{\Sig_n}$ of $n$ excitations are defined through 
the corresponding projectors $\Sig_0 \equiv \prod_{j}^N \sig_{gg}^j$,
$\Sig_1 \equiv \sum_{j}^N \sig_{rr}^j \prod_{i \neq j}^N \sig_{gg}^j$,
etc. Obviously $\expv{n} = \sum_n n \, p_n$. 
To quantify the probability distribution of Rydberg excitations,
we use the Mandel $Q$ parameter \cite{MandelQ}
\begin{equation}
Q \equiv \frac{\expv{n^2} - \expv{n}^2}{\expv{n}} - 1 ,
\end{equation}
where $\expv{n^2} = \sum_n n^2 \, p_n$. 
A Poissonian distribution $p_n =  \expv{n}^n e^{-\expv{n}}/n!$
leads to $Q=0$, while $Q < 0$ corresponds to sub-Poissonian
distribution, with $Q=-1$ attained for a definite number $n$ 
of excitations, $p_n = 1$.   

In the numerical simulations, we truncate the total Hilbert space to 
$\max n = 5$ Rydberg excitations and, upon verifying convergence, choose 
some minimum distance between the excitations, $\min|i-j| \equiv d \geq 1$,
which leads to $\dim \mathbb{H}_n^{(d)} = \binom{N-(d-1)(n-1)}{n}$,
reducing thereby significantly the computational Hilbert space.

\section{Results of simulations}

In our simulations, we use system parameters similar to those 
in recent experiments \cite{Schauss2015,Schauss2012}, i.e., we assume 
$^{87}$Rb atoms in a lattice with $a=532\:$nm excited from the ground 
state $\ket{g} \equiv \ket{5 S_{1/2},F=2,m_F=-2}$, by a two-photon process via 
a non-resonant intermediate state $\ket{e} \equiv \ket{5 P_{3/2},F=3,m_F=-3}$, 
to the Rydberg state $\ket{r} \equiv \ket{43 S_{1/2},m_J = -1/2}$ with 
the van der Waals interaction constant 
$C_6 \simeq  2\pi \times 2.45 \: \mathrm{GHz} \: \mu\mathrm{m}^6$.
The time-dependent Rabi frequency $\Om(t)$ and detuning $\de(t)$ 
of the excitation laser(s) are also chosen to have similar values 
to those in Ref.~\cite{Schauss2015}, which were optimized for preparing 
the $n=3$ Rydberg excitation ground state in a $N \sim 20$ site lattice.
We note that the precise shape of the $\Om(t)$ pulse of certain duration 
$\tau$ and the corresponding linear variation of $\de(t)$ are important 
for the quantitative characterization of the final state of the system, 
but the general conclusions of our study are qualitatively valid 
for other similar preparation strategies involving pulsed $\Om(t)$
with simultaneous monotonic increase of $\de(t)$ 
\cite{Pohl2010,Schachenmayer2010,vanBijnen2011,Vermersch2015}.

%%%%%%%%%%%%%%%%% FIGURE %%%%%%%%%%%%%%%%%%
\begin{figure}[t]
\includegraphics[width=8.7cm]{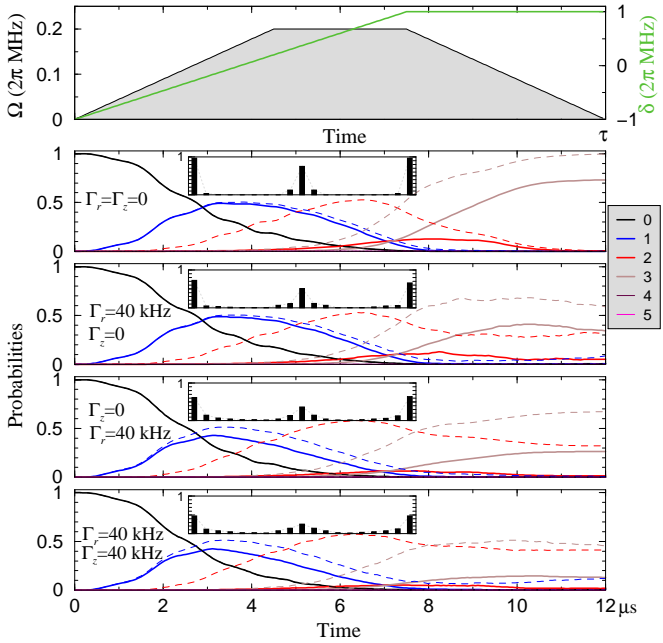}
\caption{Dynamics of the system with $N=19$ atoms initially 
in the ground state $\ket{R_0}$, subject to the time-dependent 
field with the Rabi frequency $\Om(t)$ (left vertical axis) 
and detuning $\de(t)$ (right vertical axis) shown in the top panel. 
The lower panels show the time-dependence of probabilities $p_n$ 
of $n$ Rydberg excitations (dashed lines)
and populations $P_n^{\mathrm{\min}}$ of the lowest-energy $n$-excitation 
states $\ket{R_n^{\mathrm{\min}}}$ (thick solid lines), for $n=0,1,\ldots,5$, 
in the absence or presence of atomic decay $\Ga_r$ and dephasing $\Ga_z$. 
The inset in each panel shows the corresponding spatial structure of 
the Rydberg excitation probabilities $\expv{\sig_{rr}^j}$ of atoms 
$j=1,2,\dots,N$ at the final time $\tau$. The graphs with decay 
and dephasing were obtained upon averaging over $M=150-200$ independent 
realizations (Monte-Carlo trajectories) of the numerical experiment.}  
\label{fig:N19dyn}
\end{figure}
%%%%%%%%%%%%%%%%%%%%%%%%%%%%%%%%%% 

\begin{figure}[t]
\includegraphics[width=8.5cm]{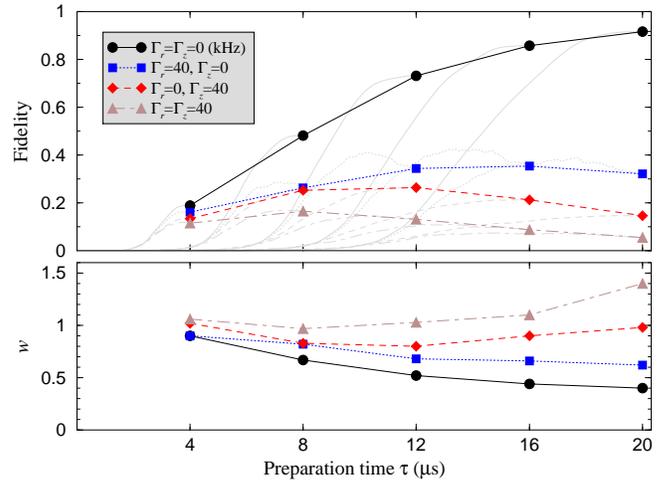}
\caption{Final fidelity $F \equiv P_3^{\mathrm{\min}}$ (top panel), and 
corresponding width $w$ (in units of $a$) of the spatial distribution 
of $\expv{\sig_{rr}^j}$ at the lattice center $j_c = 10$ (bottom panel),
versus the preparation time $\tau$, in the $N=19$ site lattice 
(as in Fig.~\ref{fig:N19dyn}) for various values of relaxation constants 
$\Ga_r,\Ga_z$ (see the legend).}  
\label{fig:N19Fw}
\end{figure}

In Fig.~\ref{fig:N19dyn} we show the dynamics of a representative 
system of $N=19$ atoms subject to an appropriate laser pulse (top panel) 
of duration $\tau = 12\:\mu$s leading, in the unitary regime 
of $\Gamma_{r} = \Gamma_{z} =0$, to the final probability $p_{3} > 0.99$ 
of $n=3$ Rydberg excitations and quite large population 
$P_3^{\mathrm{\min}} \simeq 0.73$ of the target ground state. 
This preparation time $\tau$ is significantly longer than in 
the experiment \cite{Schauss2015} with $\tau \simeq 4\:\mu$s, 
but obtaining the target ground state $\ket{R_3^{\mathrm{\min}}}$ 
with higher probability of $P_3^{\mathrm{\min}} > 0.9$ would
require even slower preparation with $\tau \gtrsim 20 \:\mu$s
(see Fig.~\ref{fig:N19Fw}). If we now add realistic decay and/or 
dephasing of the atoms, the population $P_3^{\mathrm{\min}}$ of the 
target final state would considerably decrease, see Fig.~\ref{fig:N19dyn}. 
Simultaneously, the spatial distribution of Rydberg excitations, 
while still retaining order imposed by the open boundary conditions 
\cite{Ates2012Ji2013,DPMHMF2013,Petrosyan2013}, will resemble 
perfect crystal even less.

%%%%%%%%%%%%%%%%% FIGURE %%%%%%%%%%%%%%%%%%
%%%%%%%%%%%%%%%%%%%%%%%%%%%%%%%%%%%

In Fig.~\ref{fig:N19Fw} (top panel), we show the fidelity 
$F \equiv P_3^{\mathrm{\min}}$ of attaining the target crystalline 
state of $n=3$ Rydberg excitations in the $N=19$ site lattice, for 
unitary ($\Ga_{r}=\Ga_{z}=0$) and dissipative ($\Ga_{r,z}\neq 0$) cases,
as a function of the preparation time $\tau$ [varying $\tau$ means
rescaling by the same amount the time-dependence of both $\Om(t)$ 
and $\de(t)$, see Fig.~\ref{fig:N19dyn} (top panel)]. 
We observe that, in all cases, the fidelity is rather low, 
$F \lesssim 0.2$ for $\tau = 4\:\mu$s \cite{Schauss2015}.
The resulting spatial distribution of Rydberg excitation 
probabilities $\expv{\sig_{rr}^j}$ is also very similar 
in all cases of $\Ga_r,\Ga_z$. 

In order to obtain a suitable measure for crystalline order 
in the finite system, we fit the central peak of $\expv{\sig_{rr}^j}$ 
in the vicinity of $j_c = (N+1)/2$ with a Gaussian 
$\expv{\sig_{rr}^j} \simeq A \exp [- (j-j_c)^2/ (2 w^2) ]$ and 
extract its width $w$ shown in Fig.~\ref{fig:N19Fw} (bottom panel). 
For $\tau = 4\:\mu$s we obtain $w \simeq 1a$ for all cases, which corroborates 
the above assertion that the final spatial density distribution of Rydberg 
excitation $\expv{\sig_{rr}^j}$ with or without relaxations is nearly 
indistinguishable. Apparently, when the preparation time $\tau$ is too short,
the system cannot adiabatically follow the sequence of the lowest-lying 
states $\ket{R_0} \to \ket{R_1} \to \ket{R_2^{\min}} \to \ket{R_{3}^{\min}}$ 
and the final state has a significant admixture of the higher lying 
states which are still, however, mostly within the $n=3$ excitation 
subspace due to the large energy gap with the manifold of higher $n$ states. 

%%%%%%%%%%%%%%%%% FIGURE %%%%%%%%%%%%%%%%%%
\begin{figure}[b]
\includegraphics[width=8cm]{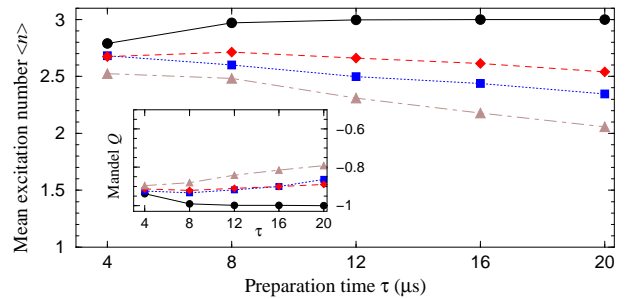}
\caption{Mean number of Rydberg excitations $\expv{n}$ (main panel)
and the corresponding Mandel $Q$ parameter (inset), versus the preparation 
time $\tau$, for the $N=19$ site lattice with various $\Ga_r,\Ga_z$;
all parameters and correspondence of symbols/lines 
are the same as in Fig.~\ref{fig:N19Fw}.}
\label{fig:N19nQ}
\end{figure}
%%%%%%%%%%%%%%%%%%%%%%%%%%%%%%%%%%%

Increasing the preparation time $\tau$ obviously improves the adiabaticity 
of the unitary evolution attested by the monotonous increase of the final 
fidelity $F$ of the target state and the decrease of $w$, as seen in 
Fig.~\ref{fig:N19Fw}. But when we add realistic decay and dephasing, 
the advantages of slower preparation largely disappear, because 
the relaxation processes acting for longer time induce more 
decoherence and deplete the crystalline ground state of the system.
Remarkably, the mean number of Rydberg excitations and the corresponding
statistics characterized by the highly sub-Poissonian Mandel $Q$ parameter, 
are considerably less susceptible to relaxations, 
as shown in Fig.~\ref{fig:N19nQ}. Note that if, 
for the given parameters of the system, we view the final spatial 
distribution of Rydberg excitations $\expv{\sig_{rr}^j}$ with 
the corresponding finite resolution $w$, it would appear nearly 
indistinguishable from the crystalline state. 

%%%%%%%%%%%%%%%%% FIGURE %%%%%%%%%%%%%%%%%%
\begin{figure}[t]
\includegraphics[width=8.5cm]{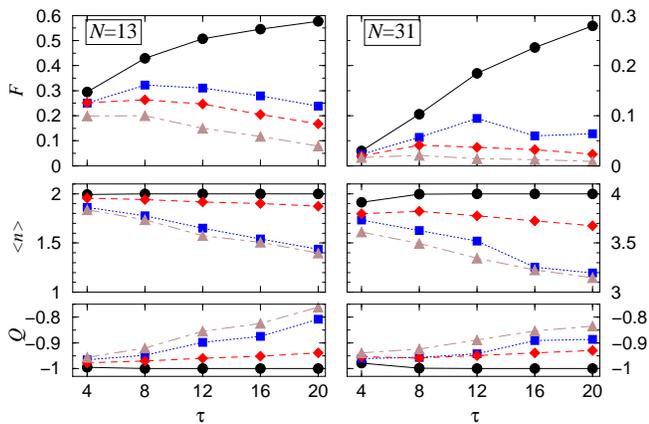}
\caption{Fidelities $F \equiv P_2^{\mathrm{\min}}$ (top left) 
and $F \equiv P_4^{\mathrm{\min}}$ (top right), 
mean number of excitations $\expv{n}$ (middle)
and the corresponding $Q$ parameters (bottom panels), 
versus the preparation time $\tau$, 
for the  lattice of $N=13$ (left column) and $N=31$ (right column) atoms.
Notice the difference in the vertical axes scales.  
All parameters and correspondence of symbols/lines 
are the same as in Figs.~\ref{fig:N19Fw} and \ref{fig:N19nQ}.}
\label{fig:N1331FnQ}
\end{figure}
%%%%%%%%%%%%%%%%%%%%%%%%%%%%%%%%%%%

As promised above, in Fig.~\ref{fig:N1331FnQ} we show the final fidelities,
together with the mean excitation numbers and the corresponding $Q$ parameters,
for a shorter lattice of $N =13$ sites that can accommodate $n=2$ excitations, 
and for a longer lattice of $N =31$ sites with up to $n=4$ excitations, 
for the same parameters as in the previous figures. 
The smaller preparation fidelity of the longer $n=4$ crystal 
is to be expected, as the system has to adiabatically follow more 
(avoided) level crossings, the last of which in the vicinity 
of $\delta_{3\to4}$ has very small level repulsion $\pm \Omega_{3\to4}$. 
What is more surprising, however, is that, for the same duration $\tau$ 
of the process, the final population of the target double excitation 
state $\ket{R_{2}^{\min}}$ in the shorter lattice is smaller than the 
final population of the target triple-excitation state $\ket{R_{3}^{\min}}$ 
in an appropriately longer lattice (cf. Fig.~\ref{fig:N19Fw}). 
%One potential reason for this counter-intuitive result might 
%be that in an $N=13$ site lattice, the ground state $\ket{R_2^{\min}}$ with 
%two excitations at the opposite ends is separated from the lowest excited 
%states by a smaller energy gap, but this explanation does not hold to the 
%argument that attaining the $n=3$ crystalline state $\ket{R_3^{\min}}$ in 
%a longer lattice still has to proceed via the $\ket{R_2^{\min}}$ state. 
This counter-intuitive behavior can be understood 
from the following consideration: 
Due to the small effective coupling $\Om_{1 \to 2} = 2 \Om/ \sqrt{N}$ 
between states $\ket{R_1}$ and $\ket{R_2^{\min}}$, and the resulting 
small level repulsion in the vicinity of $\delta_{1 \to 2}$, 
only a fraction of population of state $\ket{R_{1}}$ is 
adiabatically converted into the population of the $n=2$ ground 
state $\ket{R_{2}^{\min}}$. On the other hand, the $n=3$ crystalline 
ground state $\ket{R_3^{\min}}$ can be populated not only from 
state $\ket{R_2^{\min}}$, but also from other higher-energy 
double-excitation states $\ket{R_{2}}$ (see Fig.~\ref{fig:N19dyn}). 
Thus, much of the population remaining in $\ket{R_{1}}$ after 
its (avoided) level crossing with $\ket{R_2^{\min}}$ is transferred 
to the other double-excitation states $\ket{R_{2}}$, two of which, 
$\ket{r_1 g \ldots gr_{(N+1)/2} g \ldots  gg}$ and 
$\ket{gg \ldots gr_{(N+1)/2} g \ldots  gr_N}$, can 
later be converted into $\ket{R_{3}^{\min}}$. 
Interestingly, this partial return of population from the higher 
energy states to the adiabatic ground state of the system leads to 
its final population which is larger than would follow from a na\"ive 
use of the independent, or sequential, level crossing approximation 
\cite{Brundobler1993}.

\section{Discussion and conclusions} 

To summarize, we have shown that preparing small crystals of 
merely two to four Rydberg excitations in a lattice gas of 
several tens of atoms using the adiabatic protocol 
\cite{Pohl2010,Schachenmayer2010,vanBijnen2011} requires 
exceedingly long times to effect a slow-enough change of detuning 
of the laser field irradiating the atoms. Then, under typical 
experimental conditions \cite{Schauss2012,Schauss2015}, the relaxation
processes affecting the atoms during such long preparation times
cause multiple transitions between the diabatic energy levels and 
deplete the adiabatic crystalline ground state of the system. The
resulting mixed final state is then essentially a steady-state of 
the dissipative system \cite{DPMHMF2013,Ates2012Ji2013} subject 
to a uniform driving laser with the same final detuning. 

%The target crystalline state of a system of atoms in a one-dimensional
%lattice corresponds to a classical state with equidistant separation between 
%the Rydberg-excited atoms. In an experiment, the fidelity of preparation of a 
%of such an $n$-excitation state is then the probability of simultaneously
%detecting $n$ Rydberg atoms at the prescribed positions, which is obtained 
%from many repetitions of the preparation and site-resolved measurement cycles.
%
In an experiment, the fidelity of preparation of a target crystalline 
state of $n$ excitations of atoms in a one-dimensional lattice is
the probability of simultaneously detecting $n$ Rydberg atoms at 
equidistant positions, which is obtained from many repetitions of 
the preparation and site-resolved measurement cycles.
In our somewhat idealized treatment, we assumed a perfect lattice
of atoms with unity filling of each site, and neglected the motion of 
Rydberg-excited atoms and their detection errors. Clearly, the initial 
site-occupation defects of the trapped ground-state atoms, the motion 
and loss of the untrapped Rydberg-excited atoms, as well as finite 
detection efficiency will result in further reduction of the measured 
preparation fidelities of crystalline phases of Rydberg excitations. 

While these results may appear discouraging for the prospects of 
attaining sizable crystals of Rydberg excitations in laser-driven 
atomic media, our simulations of dissipative dynamics of the system 
still reveal spatial ordering of Rydberg excitations and highly 
sub-Poissonian probability distribution of the excitation number, 
which should not be very sensitive to site-occupation defects. 
We note that similar and even larger structures can be obtained, 
or ``grown'', perhaps more efficiently if, instead of slowly changing 
the detuning of the laser uniformly irradiating a chain of atoms initially 
in the ground state, one sweeps the laser beam with a fixed frequency 
form the one end of the chain to the other \cite{Hoening2013}.

We hope that our analysis and result are both stimulating and important 
for the general field of simulating interacting, dissipative many-body 
systems and imitating their various phases with Rydberg atoms.

\begin{acknowledgments}
This work was supported by the H2020 FET Proactive project RySQ (D.P. and K.M.),
the Villum Foundation (K.M.), and the DFG through SFB-TR49 (M.F.).
D.P. is grateful to the University of Kaiserslautern for hospitality and to 
the Alexander von Humboldt Foundation for support during his stay in Germany.
\end{acknowledgments}

\end{document}